# Co-link analysis as a monitoring tool: A webometric use case to map the web relationships of research projects

Jonathan Dudek[1], David G. Pina[2] and Rodrigo Costas[3]

[1] *j.dudek@cwts.leidenuniv.nl*
Centre for Science and Technology Studies (CWTS), Leiden University, Leiden (The Netherlands)
Delft Institute of Applied Mathematics, Delft University of Technology, Delft (The Netherlands)

[2] *David.Pina@ec.europa.eu*
European Research Executive Agency, European Commission, Brussels (Belgium)

[3] *rcostas@cwts.leidenuniv.nl*
Centre for Science and Technology Studies (CWTS), Leiden University, Leiden (The Netherlands)
Centre for Research on Evaluation, Science and Technology (CREST), Stellenbosch University (South Africa)

**Abstract**
This study explores the societal embeddedness of the websites of research projects. It combines two aims: characterizing research projects based on their weblink relationships, and discovering external societal actors that relate to the projects via weblinks. The study was based on a set of 121 EU-funded research projects and their websites. Domains referring to the websites of the research projects were collected and used in visualizations of co-link relationships. These analyses revealed clusters of topical similarity among the research projects as well as among referring entities. Furthermore, a first step into unveiling potentially relevant stakeholders around research projects was made. Weblink analysis is discussed as an insightful tool for monitoring the internal and external linkages of research projects, representing a relevant application of webometric methods.

**Introduction**

Publicly funded research projects are a prominent means to develop scientific research. In the context of Horizon 2020 (H2020), the EU's Framework Programme for research and innovation from 2014 to 2020, a broad variety of projects have been funded. Given the scope, size, and duration of many of those, evaluating their impact in a timely manner has to go beyond common metrics for scientific output, such as publications. Those, as well as citations, provide relatively belated signals of the – academic – impact of projects, and usually at a time when projects have already been finished. A more general question is whether funded research projects are raising any interest or attention in society, and whether this can be captured timelier. One of the measures to explore a project's embeddedness in society can be the analysis of the *digital response* that a project's web presence generates. In this, a research project's website can serve as a point of reference to which other entities on the Internet *respond digitally* in the form of weblinks.

*Weblinks and webometrics*

The field of webometrics has produced a broad body of research studying weblinks, how they form networks, and, among a variety of other questions, how websites can be ranked and measured based on their connections (Thelwall, 2012). However, researchers have also advised caution regarding inferences based on one-sided quantitative analyses of weblinks (Thelwall, 2006). The challenge of adequately characterizing and interpreting weblinks is considerable. While weblinks can be seen as *web-'citations'*, in a way comparable to scientific citations, such equivalence can be problematic when not considering the different motivations and contexts behind weblinks (Björneborn & Ingwersen, 2004). Nonetheless, it has also been argued that weblinks may be useful for covering the phases of research that precede the formal publication of research output (Thelwall, Klitkou, Verbeek, Stuart, & Vincent, 2010).

*Using weblinks to characterize the digital response to research project websites*

Considering the conceptual caveats around weblinks, it is obvious that the incoming links attracted by a website need to be turned into processable and meaningful metrics. This can be achieved by a combination of measures and may include a focus on the origins of web references (Thelwall, 2006). Furthermore, weblinks can – mimicking citation relationships – be combined in co-citations and bibliographic couplings (Björneborn & Ingwersen, 2004). This might as well be described as "heterogenous couplings" (Costas, de Rijcke, & Marres, 2020) more broadly. As far as heterogenous couplings assume networked structures underlying science-non-science interactions, they may be used for creating networks of web relationships as well. This would account for the fact that websites are (usually) embedded in networks of other web entities and thus, provide an opportunity for contextualizing the unique *digital footprint* of e.g., a project website. At the same time, considering a 'community' of web pages can also reveal insights regarding how much it refers to itself, as well as the extent to which it is being referenced from outside.

Our study aims to apply the concept of heterogenous couplings on the characterization of the weblink relationships of research projects. We investigate the relationships between the websites of research projects based on weblinks among them, as well as weblinks coming from outside their community. Accordingly, this research in progress revisits existing webometric methods and applies them to the analysis of the websites of publicly funded research projects.

**Methodology**

Our analysis was based on 121 research projects within the *Science with and for Society* (SwafS) program, funded under H2020, representing 77% of all SwafS projects funded by the end of 2020. We selected this portfolio of projects as study case because their objectives are linked to aspects such as science communication, careers and education, promoting gender equality, responsible R&I, and research ethics and integrity, and the engagement of citizens. The type of outputs and outcomes delivered by these projects requires the search or definition of valid forms of assessing their impact and monitoring their activities beyond the typical research related metrics. All projects included had started between 2015 and 2020. At the time of this analysis, 46 (38%) of all projects had ended already, the rest was still running. Each of the projects had a dedicated project website.

*Identifying referring sources*

To identify the web sites linking to a research project's website(s), we relied upon *backlinks*. Backlinks are the links that a website receives (Björneborn & Ingwersen, 2004). Multiple commercial services crawl the Internet for such backlinks and extract them. We used Majestic, an exhaustive source for webometric research (Orduña-Malea, 2021). This service operates a large database of weblinks and is broadly used for the task of optimizing web presences.[1]

Getting the backlinks for the projects' websites from Majestic, we did not use the path (or "URL") of the homepages of the project websites, since this would only return the backlinks referring to this specific page. Instead, we looked up the *root domain* in a given URL. This expanded the set of results by returning any link received by any webpage available under the root domain.[2] For a given website this means that both backlinks to the homepage and any sub-page are found, providing a more complete picture of the embeddedness of a website. We did not collect all the backlinks, but only focused on the referring domains. This was in order to prevent results from being biased by a large number of backlinks coming from one single domain (i.e., any link relationship between a project website and a referring website was counted just once). We collected all referring domains accordingly on January 13, 2021.

---

[1] https://majestic.com/company/about
[2] For example, in the URL https://europa.eu/, the part 'europa.eu' is the root domain.

Since the SwafS projects are funded and conducted in a European context, we limited the referring domains to those originating from an EU member state or an associated country, based on the country codes coming with each referring domain or where the so-called top-level domain (TLD) indicates such a country.[3] This resulted in a final set of 4,606 referring domains. The projects in our set might have had different chances of gaining online attention, depending on the time they had already been running. We tested for a correlation between the number of days a project had been existing from its starting day until January 13, 2021, and the number of referring domains, returning a positive, moderate correlation ($r_s = .491$).

*Creating networks based on co-link relationships*

To analyze how the SwafS project websites relate to each other, we considered the overlap between referring domains and the project websites' root domains. Per project, we calculated the share of referring domains coming from other SwafS projects among all referring domains. Then, we created a matrix of projects based on the number of other projects that refer to both of them, establishing a *co-linked* relationship between them (see the graph on the left in Figure 1). In the field of bibliometrics, this is also known as a *co-citation* (Björneborn & Ingwersen, 2004; Costas et al., 2020). Next, we created the same matrix of co-linked projects, but this time using the number of *external* domains linking to more than one SwafS project. Finally, we connected the referring, external domains based on mutually linked project websites (see the graph on the right in Figure 1). This *co-linking* is also referred to as *bibliographic coupling* (Björneborn & Ingwersen, 2004; Costas et al., 2020).

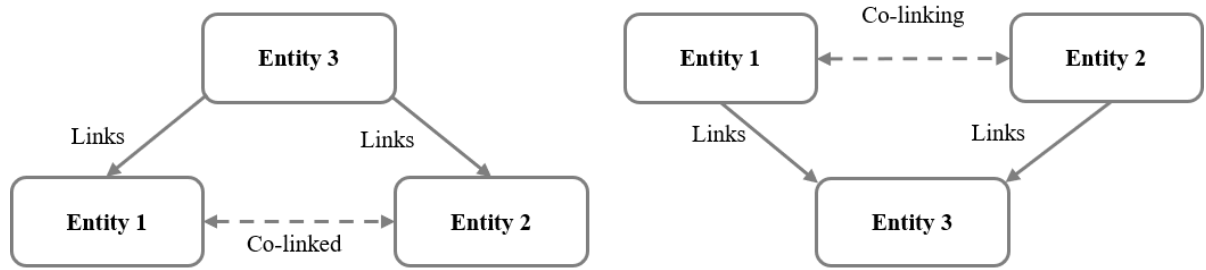

**Figure 1. Schemes for a co-linked relation (left), and a co-linking relation (right).**

In the networks with external referring domains, we did not include those with a '.com'-TLD. With 805 distinct cases found, this was the most common TLD, followed by '.eu' (424) and '.org' (390). Our reasoning was that domains with this dominant TLD could divert the focus from more interesting actors, which we assumed to rather be present with organization ('.eu', '.org') or country ('nl', 'de', etc.) related domains. For example, 90.1% of the project domains itself have an '.eu'-TLD. All networks were visualized with the VOSviewer software.[4]

**Results**

Overall, we observed a considerable weblink interconnection within the SwafS projects community. On average, 10.6% of the domains that refer to a SwafS project's website came from other SwafS projects in our set. However, the extent to which this was the case varied a lot: 24 projects were not referred to by any other project, whereas in the case of one project, 43.5% of all referring domains originated from other SwafS projects' websites. The extent of the overlap becomes also apparent when mapping the weblinks among projects based on the number of other projects that refer to them (see Figure 2). What is relevant here are different clusters of projects that are formed based on weblinks. For example, the green cluster on the right refers to projects that are aimed at (gender) equality. The blue cluster (bottom left) can be related to citizen science. In between these clusters, the red and yellow clusters include projects focused on Responsible Research and Innovation (RRI).

---

[3] The top-level domain is the URL part following the so-called second-level domain. E.g., in the URL https://europa.eu/, 'eu' is the top-level domain, and 'europa' is the second-level domain.
[4] https://www.vosviewer.com/

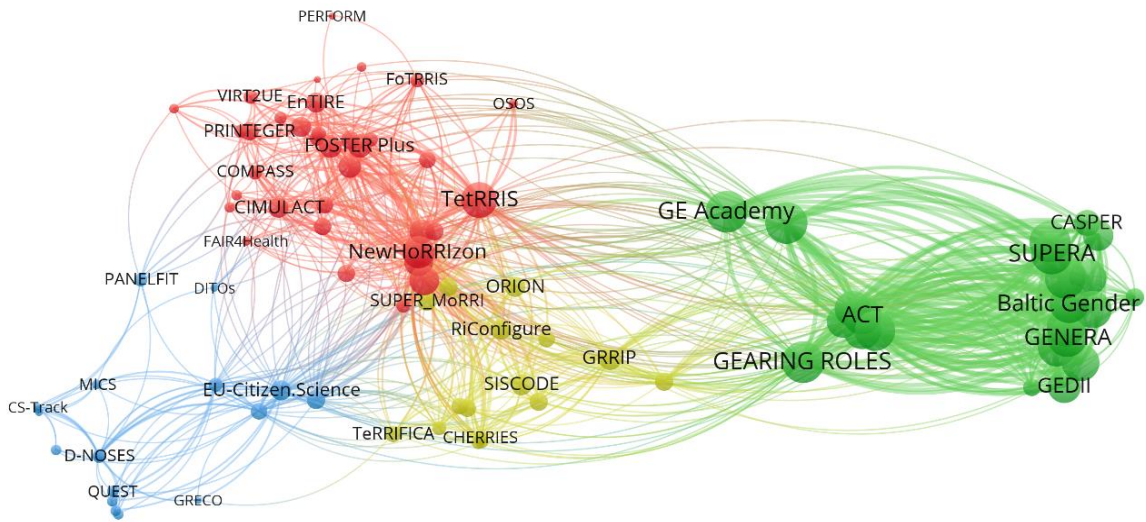

**Figure 2. VOSviewer visualization of a co-linked network of research projects based on referring other research projects.**

A similar pattern of clusters becomes visible when again looking at the co-linked network for the research projects, but this time only based on external sources of reference (Figure 3).

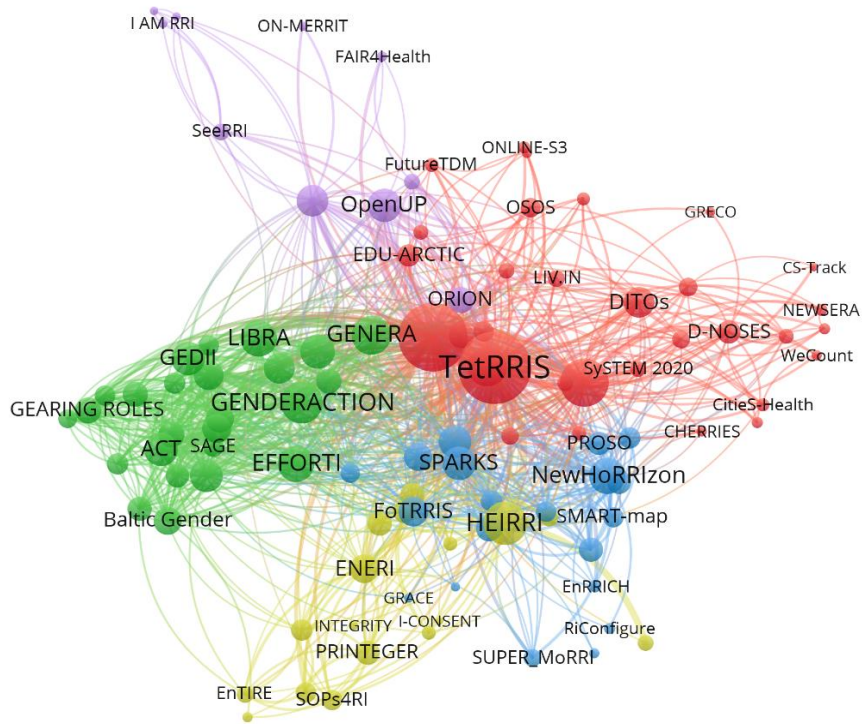

**Figure 3. VOSviewer visualization of a co-linked network of research projects based on external referring domains.**

Finally, coupling external referring domains based on concurrently cited research projects (Figure 4) revealed several clusters that reflect some of the topics also present in the cluster networks in Figures 2 and 3: For example, we again find a cluster related to gender equality (yellow), and a cluster relatable to citizen science (dark blue). The linking organizations include, among others, universities, science networks, museums, research institutes, and funding organizations.

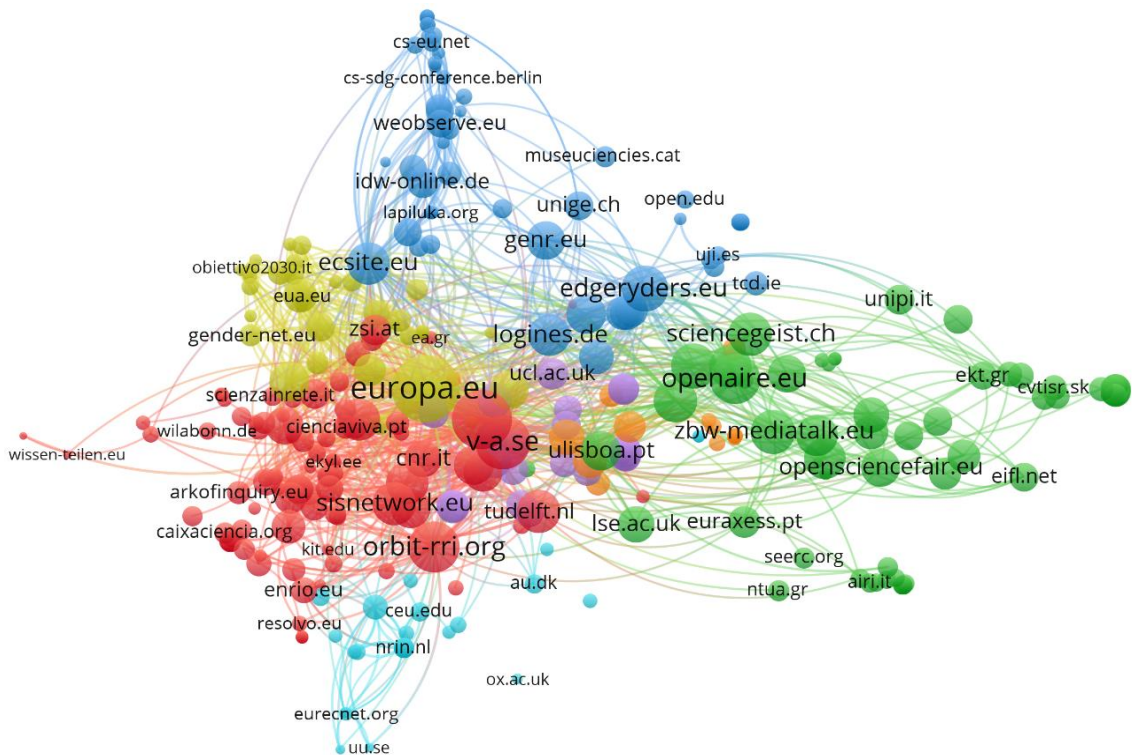

**Figure 4. VOSviewer visualization of the co-linking network of referring domains.**

**Discussion**

Our study shows how the weblink relationships of SwafS research projects capture underlying topical clusters. This could provide valuable insights into the relatedness of research projects and hint to possible synergies between them. Our proof of concept also shows the network of external actors based on the weblink connections with the projects. These external linking actors represent a unique source of relevant information on potential societal stakeholders (e.g., governments, museums, networks, etc.) who exhibit an interest in the projects. This hints to the potential value of weblink analyses to assess the 'early' societal relevance of (funded) projects, making this also very relevant for funding organizations. They could identify latent relationships existing among their funded projects at an early stage of the projects.[5]

The approach presented has relevance for both the 'community' analysis of research projects (i.e., the analysis of a set of projects from the same program or topic), and the analysis of individual projects. Combined with background information on the projects, a weblink analysis could illuminate underlying characteristics of individual projects (e.g., the relationships with other projects, or with other organizations), providing a unique tool to identify potential stakeholders relevant to each project.

---

[5] Project websites are typically created at the early stages of projects, while other outputs that could also point to project relationships – like reports or publications – generally take more time to be produced.

We conclude that the weblink analysis of scientific projects represents a strong monitoring tool that is able to characterize the internal and external linkages resulting from funded projects. This adds to the variety of applications of webometrics as well, and, more specifically, to existing research on science-related web presences (Orduña-Malea, 2021).

It should be noted that the approach presented is only a *snapshot in time*. The Internet is dynamic, and the numbers of referring domains can change quickly over time. Furthermore, research projects may have different requirements and ambitions regarding their online presence, which is connected to the numbers of links they may possibly receive. This requires caution especially when making inferences for individual cases.

We plan to further refine the approach described, including the analysis and classification of the referring sources to better grasp the underlying structures. In addition to that, insights could be obtained by breaking down results according to the countries of origin of referring sources and by connecting them to the origins of project participants.

We are also going to investigate how the approach described plays out in a different context – namely, on Twitter. For that, we aim to study connections between the Twitter accounts of the research projects and other Twitter accounts. This could eventually lead us to uncovering the "communities of attention" (Haustein, Bowman, & Costas, 2015) around research projects and thus, result in a more complete picture of the societal linkages to research projects.

## Acknowledgements

This research is partially funded by the South African DST-NRF Centre of Excellence in Scientometrics and Science, Technology and Innovation Policy (SciSTIP). We would like to thank Enrique Orduña-Malea for helpful comments on the draft of this research in progress paper.

## Disclaimer

All views expressed in this article are strictly those of the authors and may in no circumstances be regarded as an official position of the European Research Executive Agency or the European Commission.